\documentstyle[aps]{revtex}
\begin{document}

\def\eeq{\end{equation}}
\def\beq{\begin{equation}}
\def\bea{\begin{eqnarray}}
\def\eea{\end{eqnarray}}


\title{Damage spreading in the Bak-Sneppen model: \\
Sensitivity to the initial conditions and equilibration dynamics}
\author{Ugur Tirnakli\thanks{e-mail: tirnakli@sci.ege.edu.tr}} 
\address{Department of Physics,
Faculty of Science, Ege University, 35100 Izmir, Turkey}
\author{Marcelo L. Lyra} \address{Departamento de F\'{\i}sica,
Universidade Federal de Alagoas, Macei\'o-AL, Brazil}

\date{\today}
\maketitle

\begin{abstract}
The short-time and long-time dynamics of the Bak-Sneppen model
of biological evolution are investigated using the
damage spreading technique. By defining a proper Hamming distance
measure, we are able to make it exhibit an initial power-law growth
which, for finite size systems, is
followed by a decay towards equilibrium. In this sense,
the dynamics of self-organized critical states is shown to be
similar to the one observed
at the usual critical point of continuous phase-transitions and at the onset
of chaos of non-linear low-dimensional dynamical maps.
The transient, pre-asymptotic and asymptotic exponential relaxation
of the Hamming distance between two initially
uncorrelated equilibrium configurations is also shown to be fit
within a single mathematical framework.

\end{abstract}


\vspace{1cm}

\section{Introduction}

The concept of self-organized criticality was introduced
to describe the tendency of large dynamical systems to organize
themselves in an out of equilibrium state exhibiting spatio-temporal
complexity\cite{bak}. Such complexity is reflected by the presence
of correlations between events separated in space and time over a
wide range of length and time scales\cite{zhang,fogedby}. The emergence of
spatio-temporal complexity in dynamical systems driven by rules based
on extremal principles is in the origin of the widespread occurrence
in nature of fractal structures\cite{mandelbrot}, noise with $1/f$
power spectrum and punctuated equilibrium\cite{paczuski}.

The simplest mathematical model of a driven dynamical system,
which exhibits self-organized criticality, was introduced by Bak
and Sneppen as a toy model of biological evolution of an ecology
of interacting species\cite{sneppen}. In this model random numbers
$f_i$ (fitness), uniformly distributed on the interval [0,1], are
assigned to the sites of a $d$-dimensional lattice. At each time step
the site with the smaller random number is located and new random
numbers from a uniform distribution are assigned to that site
and also to its first neighbors.

The above model system achieves a statistically stationary state
in which the density of random numbers in the system vanishes for
$f<f_c$ and is uniform above $f_c$, with $f_c =0.667$ \cite{sneppen} in
a chain and $f_c =0.38$ in a square lattice geometry \cite{paczuski}.
Once the stationary state is achieved the system exhibits punctuated
equilibrium as real biology\cite{gould}, characterized by intermittent
co-evolutionary avalanches of all sizes. The complexity of
this stationary regime can be revealed by the show up of
spatio-temporal power-law distribution of events. For example,
the distribution $C(x)$ of the distance $x$ between two
subsequent extremal sites scales as $C(x)\propto x^{-\pi}$,
with $\pi = 3.23$ in one dimension \cite{paczuski}. Further the temporal
long-range correlations can be also observed in the distribution
of first return times $P(t)$ which scales as $t^{-\tau_{first}}$, with
$\tau_{first} = 1.58$ in $1d$ and $\tau_{first} = 1.28$
in $2d$ \cite{paczuski}.

The propagation of local perturbations in the self-organized state
of Bak-Sneppen model was recently investigated using the
damage spreading algorithm\cite{tamarit,gleiser,cafiero,cafiero2,valleriani}.
By measuring how the difference between two initially close configurations
evolves in time under the same noise, it was shown that an initial power-law
divergence sets up before the distance saturates
in a finite size-dependent plateau. This slow short-time power-law dynamics 
is characteristics of systems poised at critical states such as
at the critical point of second-order phase transitions\cite{janssen} and the
onset of chaos in non-linear dynamical
systems\cite{tpz,lyra1,lyra,bjp,circular,asymmetric,fulvio,singlesite,henon}.

However, usual critical phenomena exhibit two dynamical
regimes characterized by distinct critical exponents.
The first one is related to a short-time slow dynamics, which governs
the power-law increase of the order parameter when its value in an
initial non-equilibrium configuration is small and non-zero\cite{janssen}.
In finite systems this power-law increase persists up to a characteristic
time, after which the order parameter slowly decays towards its
equilibrium value.

The same dynamical regimes have been identified in the onset of chaos
of low-dimensional non-linear dynamical maps. The distance between two
nearby orbits was shown to diverge following a power-law whose exponent
is directly related to geometric exponents characterizing
the extremal sets of the critical dynamical
attractor\cite{tpz,lyra1,lyra,bjp,circular,asymmetric,fulvio,singlesite,henon}.
On the other hand,
the long-time relaxation towards the dynamical attractor is governed
by a new exponent, which seems to be related to the fractal dimension of
the support of the dynamical attractor in phase-space\cite{moura}.

Damage spreading studies of critical phenomena have also reported two
dynamical regimes with distinct and independent exponents. Therefore, it
is natural to expect this to be the general trend of systems poised
at criticality as well as systems exhibiting
self-organized criticality such as the Bak-Sneppen model. The fact that
previous damage spreading works have not identified the long-time relaxation
towards equilibrium in the BS model just reflects the insensitivity of
the proposed measure to capture the long-time dynamics.

In this work we introduce a new Hamming distance measure that is
sensitive to the long-time dynamics of the Bak-Sneppen model. Following its
temporal evolution we are going to show that, besides displaying a power-law
short-time dynamics with the same exponent previously reported, it also
exhibits a long-time relaxation dynamics.
We will report the scaling properties of this regime and
discuss its possible relation with the geometric properties of the
dynamical attractor in phase space.

\section{Model system and numerical simulation}

In our simulations we implement the Bak-Sneppen algorithm in linear chains
with periodic boundary conditions. Random numbers $f_i$ (fitness) are
initially assigned to all sites from a uniform distribution in the
interval $[0,1]$.
The system evolution is based on the standard re-assignment of the extremal
site, i.e., the site with the minimum fitness and of its first neighbors.
In our simulations we worked with chains of up to $N=1000$ sites and
left $t_{trans}=20N$ collective time steps for the system to achieve the
statistically stationary state. Our collective time unit corresponds to
$N$ elementary time steps so that each site is going to be updated only once
in average during a collective time step.

After achieving the statistically stationary state we produce a copy of
the system's configuration. A small damage is introduced in the copy by
interchanging the position of the site with the minimum fitness with a
randomly chosen one. After this, we follow the temporal evolution of both
configurations using always the same set of random numbers to update both
replicas. After a characteristic time $\tau$, both configurations will be
composed by the same sequence of random numbers just shifted by a random
distance. Once periodic boundary conditions are used, these two
configurations are indeed indistinguishable and should be considered as
identical. In figure~1 we report the finite size scaling of the average
time needed for two initially random configurations to become identical
when updated following the Bak-Sneppen dynamics and under the same set of
random numbers. This characteristic time, measured in units of collective
time steps, scales as $\tau\propto N^{\phi}$, with $\phi =1.46(4)$.
We would like to stress that, within the error bars, this is the same scaling
exponent obtained by the size dependence of the average self-organization time
to reach the critical attractor [$D-1 = 1.43(1)$]\cite{paczuski}
and is independent of the initial state of the system.

In order to accomplish our task, we introduce a new Hamming
distance $D(t)$ defined at each time $t$ as the smallest among the $N$
possible values of $D_j(t)$ ($j=1,2,...N$) defined as
\begin{equation}
D_{j}(t) = \left< \frac{1}{N}\sum_{i=1}^{N}|f_{i}^{1}-f_{i+j}^{2}|\right>~~.
\end{equation}
where $\langle ... \rangle$ represents configurational average
over distinct runs. Those runs at which the Hamming
distance vanishes are not considered in the averaging process
and therefore our simulations are limited in size by the characteristic
time $\tau$.
Notice that $D_0(t)$ corresponds to the measure used in previous damage
spreading studies of Bak-Sneppen
model\cite{tamarit,gleiser,cafiero,cafiero2,valleriani}.
After the initial power-law growth $D_0(t)\propto t^{\alpha}$ with
$\alpha = 0.32$, it saturates at a constant value
which can be shown to be $1/3$ of the width of the fitness
distribution in equilibrium\cite{valleriani}.
This saturation is due to the
fact that by just measuring $D_0(t)$ one cannot identify that both replicas
have converged to the same random sequence configuration.

The procedure proposed here is to measure the Hamming distance as the
minimum distance between the replicas taking in account all possible shifts
between them. In respect to the short-time dynamics it is equivalent to
measuring $D_0(t)$ itself and shall give the same power-law exponent for the
initial Hamming distance growth. However, after achieving a maximum,
it must decrease to vanishing small values as both replicas converge to
the same sequence, capturing then the long-time dynamics.

In figure~2 we show our results for $D(t)$ for chains up to $N=1000$ sites.
As larger chain sizes are considered the initial power-law regime extends
for longer periods. The dashed line corresponds to the
power law $D(t)\propto t^{\alpha}$ with $\alpha = 0.32$ in full agreement
with the value previously reported in the literature\cite{tamarit}.
After a size dependent characteristic time, it reaches a maximum and starts
to monotonically decrease. However, much longer runs than these presently
reported would be needed to precisely estimate the scaling behavior of
the Hamming distance decay.

In order to investigate in detail the long-time dynamical regime, it is
computationally more efficient to use a slightly different approach.
We start with two uncorrelated equilibrium configurations and follow
the time evolution of the distance between them when the same set of
random numbers is used. In this way, the initial growth is absent and
the relaxation regime can be observed using shorter runs.
Usually the critical power-law relaxation dynamics takes place
after some transient time. This typical behavior means that
the Hamming distance at criticality satisfies the non-linear
differential equation
\begin{equation}
\frac{dD(t)}{dt}= -\lambda D^q
\end{equation}
whose solution is
\begin{equation}
D(t) = D(0)/[1+\lambda(q-1)t]^{1/(q-1)}~~.
\end{equation}
and reproduces the above trend. In numerical simulations, the
critical relaxation dynamics is always followed by an exponential
relaxation due to finite size effects. A pure exponential relaxation
implies that $D(t)$ must asymptotically obey a linear differential equation.
Therefore in order to consider simultaneously the critical and exponential
relaxation we consider $D(t)$ to obey a more general non-linear differential
equation such as
\begin{equation}
\frac{dD(t)}{dt}= -(\lambda -\mu )D^q - \mu D ~~.
\end{equation}
In the above equation, $\mu$ represents the finite size correction and 
vanishes in the thermodynamic limit. The general solution of the above 
differential equation is in the form
\begin{equation}
D(t) = D(0)/[1-(\gamma /\mu)+(\gamma /\mu)e^{(q-1)\mu t}]^{1/(q-1)}~~.
\end{equation}
where $\gamma = \mu-(\mu - \lambda)/D_0^{(1-q)}$. It is worth to mention that 
similar equations have also been used to describe experimental data on the 
re-association in folded proteins\cite{tsallis}, quantitative 
linguistics\cite{montemurro} and fluxes of cosmic rays \cite{cosmicrays} 
within the framework of nonextensive thermostatistics \cite{tsallis2,tsallis3}.
Indeed, the above form is expected to generally describe the
sensitivity to initial conditions of systems at the vicinity of
critical states.

In Figure~3 we report our results for $D(t)$ starting with two uncorrelated
equilibrium configurations for several chain sizes. The average initial
distance equals to $D(0)=0.111$ which corresponds to $1/3$ of the width of
the interval $[f_c,1]$ where $f_c=0.667$ for the $1d$ Bak-Sneppen model.
The coefficient $\mu$ determines characteristic exponential decay
which sets up in the long-time regime as can be seen in the inset.
The solid lines correspond to best fits to the form of Eq.~(5).
We recall that the fitting parameters govern the typical behavior at
distinct time regimes.
In the table we summarize the parameters obtained. Notice that $\mu$ decreases
monotonically when chain size is increased, once the exponential relaxation
is postponed for large chains. Also $\gamma$ decreases with $N$ so that
the transient period diverges in the thermodynamic limit. However,
$\mu \ll \gamma$ and therefore the pre-asymptotic decay can be
observed just at intermediate times satisfying
$1/\gamma \ll t \ll 1/\mu$.

In figure 4 we show the size dependence of $\mu$ and $\gamma$. 
Once $1/\mu$ determines the time scale needed to finite size effects become 
relevant, it shall present the same scaling behavior as the average time 
$\tau$ for both replicas to become identical.
We found $1/\mu\propto N^{1.49(4)}$, which is consistent with
the above conjecture. On the other hand, $1/\gamma$ defines the time scale 
when the crossover between the transient and the pre-asymptotic regime takes 
place. It is governed by a distinct exponent $1/\gamma\propto N^{\kappa}$, 
with $\kappa = 1.11(3)$ and is directly related to the average time needed 
to update all active sites in the initial avalanche\cite{paczuski}.

For the system sizes simulated, the intermediate regime satisfying the 
condition $1/\gamma \ll t \ll 1/\mu$ is achieved only in a narrow time interval 
which grows slowly as the chain size is increased. This feature introduces 
a large uncertainty in the estimates of the non-linear exponent $q$, 
particularly the ones from small chain data. The best fitting values for 
$q$ were found to be much larger than one, specially the more confident 
estimate from chain size $N=1000$. This may indicate a very slow logarithmic 
pre-asymptotic decay of the Hamming distance.

At this point it is worth to explore the similarities between the
damage spreading analysis of the dynamical properties of extended
critical systems and recent results concerning the sensitivity to
initial conditions of low-dimensional dissipative maps poised at
criticality\cite{tpz,lyra1,lyra,bjp,circular,asymmetric,fulvio,singlesite,henon}.
More precisely, the numerical analysis of such systems (i.e., logistic
map\cite{tpz}, logistic-like maps\cite{lyra1}, circular maps\cite{circular},
asymmetric logistic maps\cite{asymmetric}, single-site map\cite{singlesite},
Henon map\cite{henon}) has shown that, at critical points such as the
chaos threshold, tangent bifurcations etc, the standard type of the 
sensitivity to the initial conditions given by the sensitivity function

\beq
\xi (t) \equiv \lim_{\Delta x(0)\rightarrow 0} \frac{\Delta x(t)}
{\Delta x(0)}= \exp\left(\lambda_1 t\right)\;
\eeq
(where $\lambda_1$ is the standard Lyapunov exponent) can be replaced by a
power-law type of sensitivity

\beq
\xi(t)=\left[1+(1-q)\lambda_q t\right]^{1/(1-q)}\; ,
\eeq
which indicates a weak sensitivity to the initial conditions. Here,
$\lambda_q$ is the generalized Lyapunov exponent, and satisfies the 
generalized Pesin equality $K_q=\lambda_q$ if $\lambda_q >0$ and $K_q=0$ 
otherwise, where $K_q$ is the generalized Kolmogorov-Sinai entropy \cite{tpz}.

Now we are in a position to discuss the connection between this scenario
and the BS model. Indeed, the BS model belongs to another class of
dynamical systems, namely high-dimensional dissipative systems, for which this
scenario is also needed to be tested. As a matter of fact, the BS model
evolves spontaneously (without a tuning parameter) towards a critical state,
whereas a one-dimensional map approaches to a critical point (e.g. chaos
threshold point) with a fine tuning of a map parameter. Moreover, in both
cases, the standard Lyapunov exponent would be zero. Therefore, the
properties of the sensitivity to the initial conditions of the BS model are
expected to be similar to those of the low-dimensional dissipative maps at
the edge of chaos. Sensitivity to the initial conditions, which is reflected
by the sensitivity function for the low-dimensional dissipative maps,
can be conveyed by the damage spreading (and Hamming distance) for the BS
model. It is clear from ref.\cite{tamarit} and from our results using a
different definition of the Hamming distance that a power-law spread of damage
emerges for the BS model in the short-time dynamics regime. Then, one can
easily determine the proper $q$ value of the BS model as $q^*\simeq -2.1$.

In what follows we discuss the picture seen from the analysis of long-time
dynamics of the BS model. Firstly, we should recall the results of
low-dimensional dissipative maps. Recently, it has been shown \cite{moura}
that, for the long-time dynamics of such dynamical systems, a new class of
$q$ values emerges. This new class appeared to be related to the equilibration
(or long-time dynamics). It is seen from the results of \cite{moura} that,
whenever the critical attractor of the dynamical system has a fractal support
(or in other words, the critical attractor is not dense), the system (e.g.,
logistic-like maps) exhibits a power-law long-time relaxation with a $q$ value 
greater than unity. On the other hand, if the critical attractor of the 
dynamical system (e.g., circular maps) is dense, then a slow logarithmic 
decay has been observed. When we look at the BS model, the situation is 
similar to that of the circular maps in the sense that the attractor in 
phase space is fully occupied in the interval $[f_c,1]$, which makes the 
critical attractor dense. The observed slow (logarithmic) pre-asymptotic 
dynamics seems to indicate that it shall take place in general system models 
whose critical dynamical attractor is dense.

\section{summary and conclusions}
In summary, we have applied the damage spreading algorithm to
study the dynamics of the Bak-Sneppen model for biological evolution which
is one of the prototypes models exhibiting self-organized criticality.
By defining a proper Hamming distance between two system's configurations we
investigated both the short-time and long-time critical dynamics.
A power-law short-time dynamics was found in agreement with previous
reports\cite{tamarit,gleiser,cafiero,cafiero2,valleriani}.
In the relaxation regime, the Hamming distance was shown to follow the
solution of a generalized non-linear equation, first proposed in the
context of nonextensive thermostatistics, which includes in a single 
expression the transient, pre-asymptotic and asymptotic exponential 
relaxation regimes.
We found that the pre-asymptotic regime is very close to a slow logarithmic
decay in a close relationship with the relaxation dynamics of non-linear
dissipative maps at the onset of chaos with a dense dynamical critical
attractor. It would be interesting to investigate the relaxation dynamics
of self-organized critical systems with a fractal dynamical attractor in
phase-space. In this case, based
on previous studies of the critical dynamics of low-dimensional deterministic
non-linear maps at the onset of chaos, one would expect a power-law 
relaxation. This point will be the subject of a future contribution.

\section{acknowledgements}
This work has been supported by the Turkish Academy of Sciences, in the 
framework of the Young Scientist Award Program (UT/TUBA-GEBIP/2001-2-20). 
M.L.L. would very much appreciate to acknowledge the hospitality of the
Physics Department of Ege University where this work was done and thanks
TUBITAK for financially supporting his visit via NATO-D Program.
Partial financial support from the Brazilian research agencies CNPq, CAPES 
and FAPEAL is also acknowledged.


\newpage

\section{Figure and Table Captions}

Figure 1 - The finite size scaling of the average time for two initially
uncorrelated random sequences to become identical when updated following
the Bak-Sneppen dynamical rules using the same set of random numbers.
The solid line corresponds to a power-law fit $\tau\propto N^{\phi}$
with $\phi = 1.46(4)~$. The number of experiments used in calculations
for different chain sizes is 100.\\

Figure 2 - The time evolution of the Hamming distance between an equilibrium
configuration and its slightly modified replica. After the initial power-law
growth $D(t)\propto t^{\alpha}$ with $\alpha = 0.32$, it monotonically
decreases as both configurations converge to the same sequence of random
numbers. The number of experiments used in calculations varies from 100
up to 500 for different chain sizes.\\

Figure 3 - The time evolution of the Hamming distance between two uncorrelated
equilibrium configurations. After the initial transient where the Hamming
distance is close to its initial value $D(0)=0.111$, $D(t)$ exhibits a
non-trivial decay followed by an exponential relaxation due to finite size
effects. The solid lines correspond to best fits to the form of Eq.~(5)
that contains the above three regimes. The fitting parameters are reported
in the table. The number of experiments used in calculations varies from 100
up to 500 for different chain sizes.\\

Figure 4 - The finite size dependence of the fitting parameters $\mu$ and
$\gamma$. The solid lines are best fits to power-laws. We found that
$1/\mu\propto N^{1.49(4)}$ and therefore it is, within the error bar,
proportional to the average time needed to both configurations coincide.
On the other hand $1/\gamma\propto N^{\kappa}$ with $\kappa = 1.11(3)$
has, within the error bar, the same scaling behavior of the average time
needed to update all sites in the initial avalanche.\\

Table Caption - The size dependence of the best fitting parameters.
Notice that $\mu$ decreases as the chain size $N$ increases, as expected.
The large values of $q$ indicate a possible slow logarithmic relaxation
in the pre-asymptotic regime as discussed in the text.\\


\vspace{1cm}

\begin{center}
{\bf Table}

\vspace{1cm}

\begin{tabular}{||c|c|c|c||} 
\hline $N$ & $\gamma$ & $\mu$ & $q$ \\ 
\hline $\;100\;$ & $\;0.088\;$ & $\;0.012\;$ & $\;8.7\;$ \\ 
\hline $\;200\;$ & $\;0.039\;$ & $\;0.0043\;$ & $\;8.7\;$ \\ 
\hline $\;500\;$ & $\;0.013\;$ & $\;0.00094\;$ & $\;10.0\;$ \\ 
\hline $\;1000\;$& $\;0.0062\;$ & $\;0.00034\;$ & $\;14.1\;$ \\ 
\hline \end{tabular} 
\end{center}

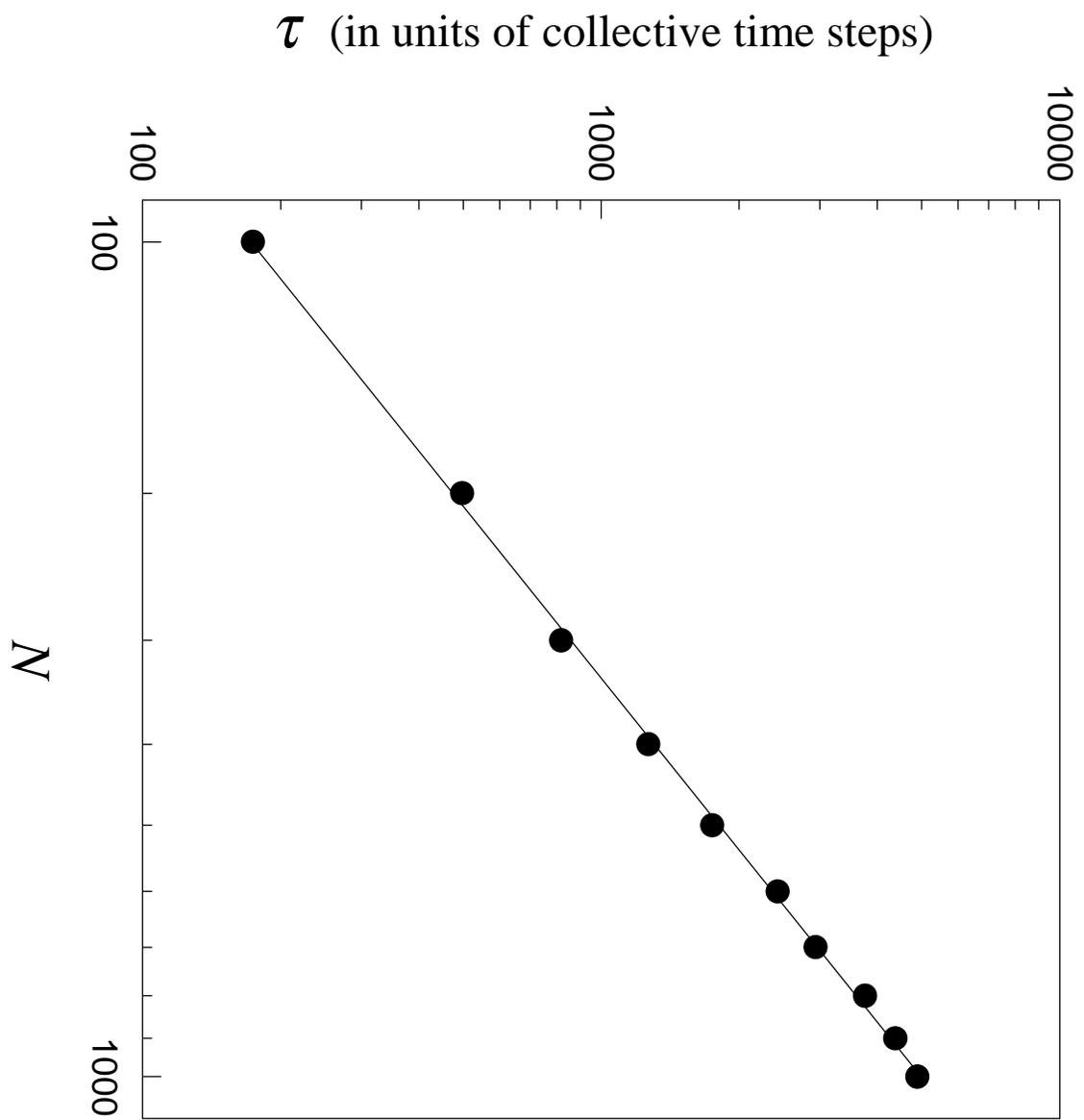
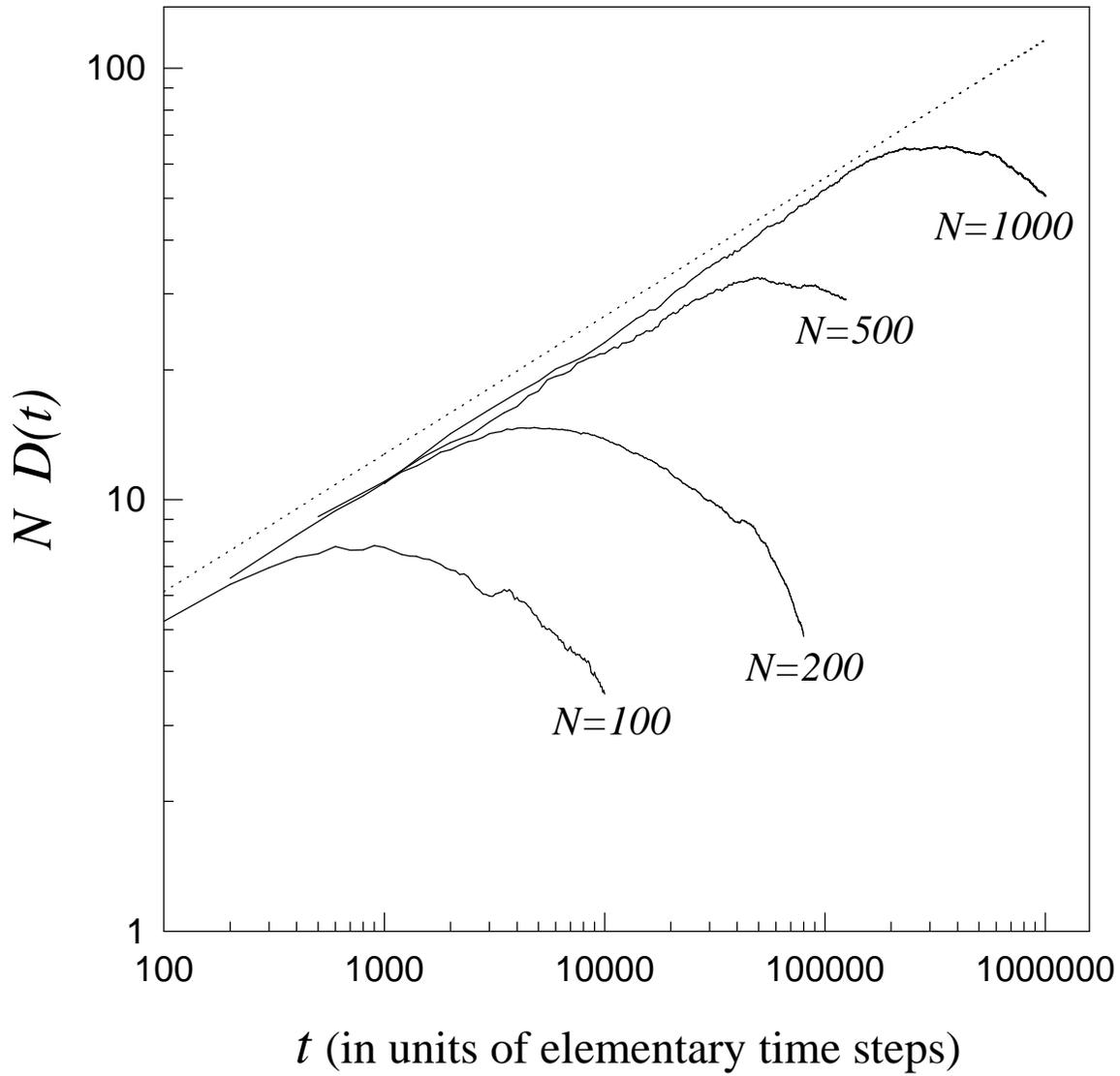
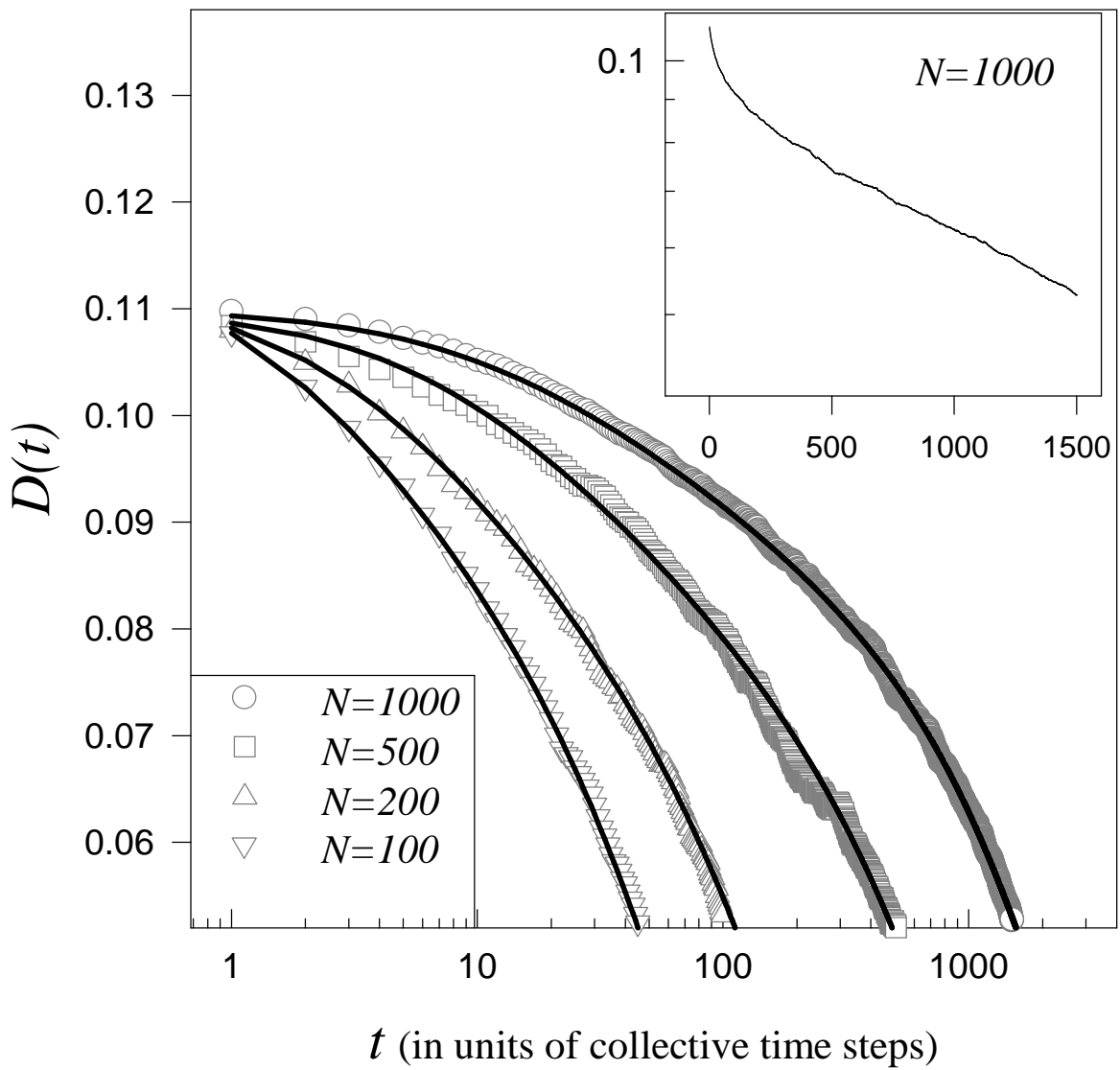
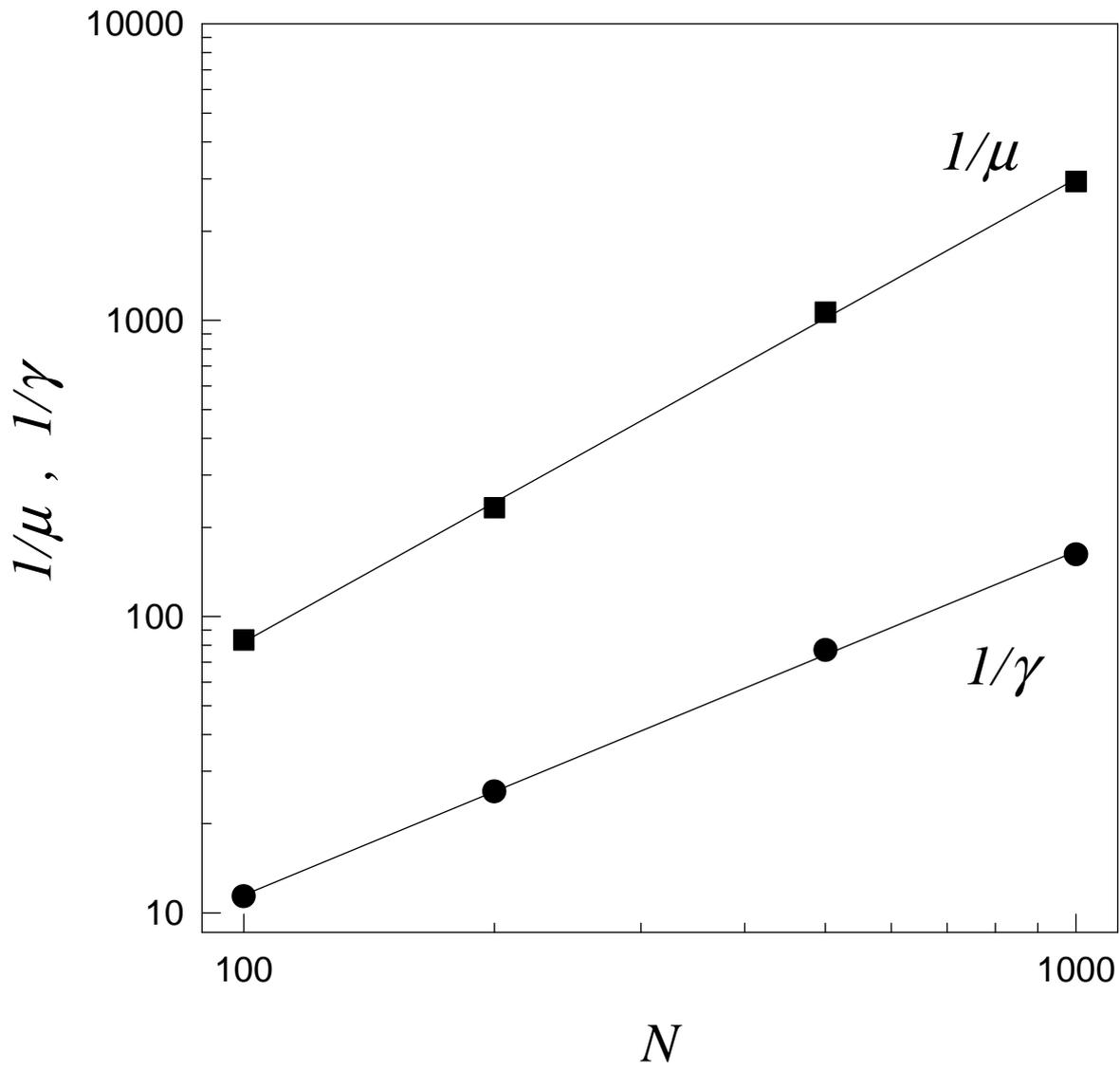
\end{document}